\documentclass[10pt,letterpaper]{article}
\usepackage{opex3}
\usepackage{graphicx}
\usepackage{cite}
\usepackage{color}
\usepackage{amsmath}
\usepackage{here}
\usepackage{epstopdf}%This line makes .eps figures into .pdf - please comment out if not required.
\newcommand{\be}{\begin{equation}}
\newcommand{\ee}{\end{equation}}
\newcommand{\bea}{\begin{eqnarray}}
\newcommand{\eea}{\end{eqnarray}}

\newcommand\pictc[5]{\begin{figure}
                       \centerline{\vspace{-1mm}
\includegraphics[width=#1\columnwidth,height=0.7\textheight,keepaspectratio]{#3}}
                       \protect\caption{\protect\label{#4} #5}\vspace{-3mm}
                    \end{figure}            }

\newcommand\pict[4][1]{\pictc{#1}{!tb}{#2}{#3}{#4}}

\begin{document}

\begin{sloppy}

\title{Unidirectional superscattering by multilayered cavities of effective radial anisotropy}

%\author{Wei Liu,$^{\ast}$ Jianfa Zhang, Bing Lei, and Haojun Hu}
\author{Wei Liu,$^{\ast}$  Bing Lei,  Jianhua Shi, and Haojun Hu}
\address{College of Optoelectronic Science and Engineering, National University of Defense
Technology, Changsha, Hunan 410073, China
\email{$^{\ast}$wei.liu.pku@gmail.com}
%$^*$Corresponding author: wei.liu.pku@gmail.com
}

\begin{abstract}% 
We achieve unidirectional forward superscattering by multilayered spherical cavities which are effectively radially anisotropic. It is demonstrated that, relying on the large effective anisotropy,  the electric and magnetic dipoles can be tuned to spectrally overlap in such cavities, which satisfies the Kerker's condition of simultaneous backward scattering suppression and forward scattering enhancement. We show such scattering pattern shaping can be obtained in both all-dielectric and plasmonic multilayered cavities, and believe that the mechanism we have revealed provides extra freedom for scattering shaping, which may play a significant role in many scattering related applications and also in optoelectronic devices made up of intrinsically anisotropic two dimensional materials.
\end{abstract}

%\maketitle

%\ocis{(240.6680) Surface plasmons; (260.2030) Dispersion}

%\bibliography{References_scattering}

\section{Introduction}

With the rapid development of the field of metamaterials and metadevices~\cite{Zheludev2012_NM,chen_review_2016}, recently the topic of scattering pattern manipulation based on the interferences of electric and artificial magnetic resonances has attracted a lot of attention (see Refs.~\cite{Liu2014_CPB,Liu2014_ultradirectional,jahani_alldielectric_2016} and references therein). The physical mechanism behind originates from the concept of Huygens source in the antenna theory~\cite{Love1976_RS,Jin2010_IEEE,Krasnok2011_Jetp} and the proposal of Kerker~\cite{Kerker1983_JOSA} to simultaneously suppress backward scattering and enhance forward scattering, which has only recently been specifically demonstrated in both all-dielectric and plasmonic nanostructures (for both dipolar and higher order modes)~\cite{Nieto2010_OE,Gomez-Medina2011_JN,Krasnok2011_Jetp,Liu2012_ACSNANO,Geffrin2012_NC,Filonov2012_APL,Krasnok2012_OE,Rolly2012_OE,Fu2013_NC,Person2013_NL,Staude2013_acsnano,
Hancu2013_NL,Vercruysse2013_NL,Liu2014_ultradirectional,Lukyanchuk2014_arXiv_Optimum}. Moreover, the scattering suppression based on resonance interferences has been extended from the backward direction to other scattering angles~\cite{Lukyanchuk2011_NM,Liu2013_OL2621,Liu2014_ultradirectional,paniagua2016generalized}, the principle of which can be applied to generalize the concept of Brewster angle~\cite{Liu2013_OL2621,paniagua2016generalized}.

In various nanostructures, usually resonances of the lowest order (dipolar resonances) are most accessible, which  can be excited with high efficiency and thus are dominant.  As a result, the achievement of the backward scattering suppression and forward scattering enhancement replying on overlapping electric dipoles (EDs) and magnetic dipoles (MDs) (the so called Kerker's condition~\cite{Kerker1983_JOSA,Alu2010_JN}) is still one of most outstanding examples of scattering pattern manipulations based on resonance interferences~\cite{Liu2014_CPB,Liu2012_ACSNANO,Geffrin2012_NC,Fu2013_NC,Person2013_NL}. It is shown that even for homogeneous dielectric spheres, the EDs and MDs can be tuned to overlap partly, leading to unidirectional forward scattering~\cite{Geffrin2012_NC,Fu2013_NC,Person2013_NL}. Nevertheless, for simple homogeneous dielectric spheres, the EDs and MDs can not be tuned to resonantly overlap (their central resonant positions do not coincide) and thus the overlapping position does not locate at the resonance centre. Consequently, though the backward scattering has been significantly suppressed, the forward scattering is not strong enough to be in the superscattering regime~\cite{Ruan2011_APL}. Recently fully resonant overlapping of EDs and MDs and thus unidirectional forward superscattering has been achieved in core-shell plasmonic nanoparticles~\cite{Liu2012_ACSNANO,Liu2014_ultradirectional}, in dielectric nanodisks~\cite{Staude2013_acsnano} and in spheroidal dielectric nanoparticles~\cite{Lukyanchuk2014_arXiv_Optimum}. For the latter two cases however, the scattering is dependent on the polarization and incident angle. Alternatively, it is recently proposed that even for homogeneous dielectric spheres, that electric radial anisotropy can be employed to enable resonant overlapping of EDs and MDs~\cite{Liu2015_OE_Ultra}. The problem is that the anisotropy required is too large to be found in naturally accessible materials.
 
In this work it is shown that multilayered cavities made up of isotropic layers can provide effective radial anisotropy that is large enough to enable resonant overlapping of EDs and MDs excited. Here we investigate the scattering of  multilayered spherical resonators and demonstrate that in both all-dielectric and plasmonic cavities the Kerker's condition can be fulfilled, and thus significant backward scattering suppression and forward superscattering can be simultaneous achieved. We note that the mechanism we have revealed is general, and should be applicable to cylindrical multilayered cavities and cavities of other shapes, to resonances of higher orders, and to other types of anisotropy~\cite{Stout2006_JOSAA_Mie,Qiu2010_IPOR_Light}. We believe that such principle provides an extra dimension of freedom for resonance control and scattering shaping, which may shed new light to many scattering related applications, and to optoelectronic devices incorporating two dimensional materials that are intrinsically highly anisotropic.

\section{Theoretical analysis on plane wave scattering by multilayered cavities of effective radial anisotropy}

%-------------------------------------------------------------------------------
\pict[0.6]{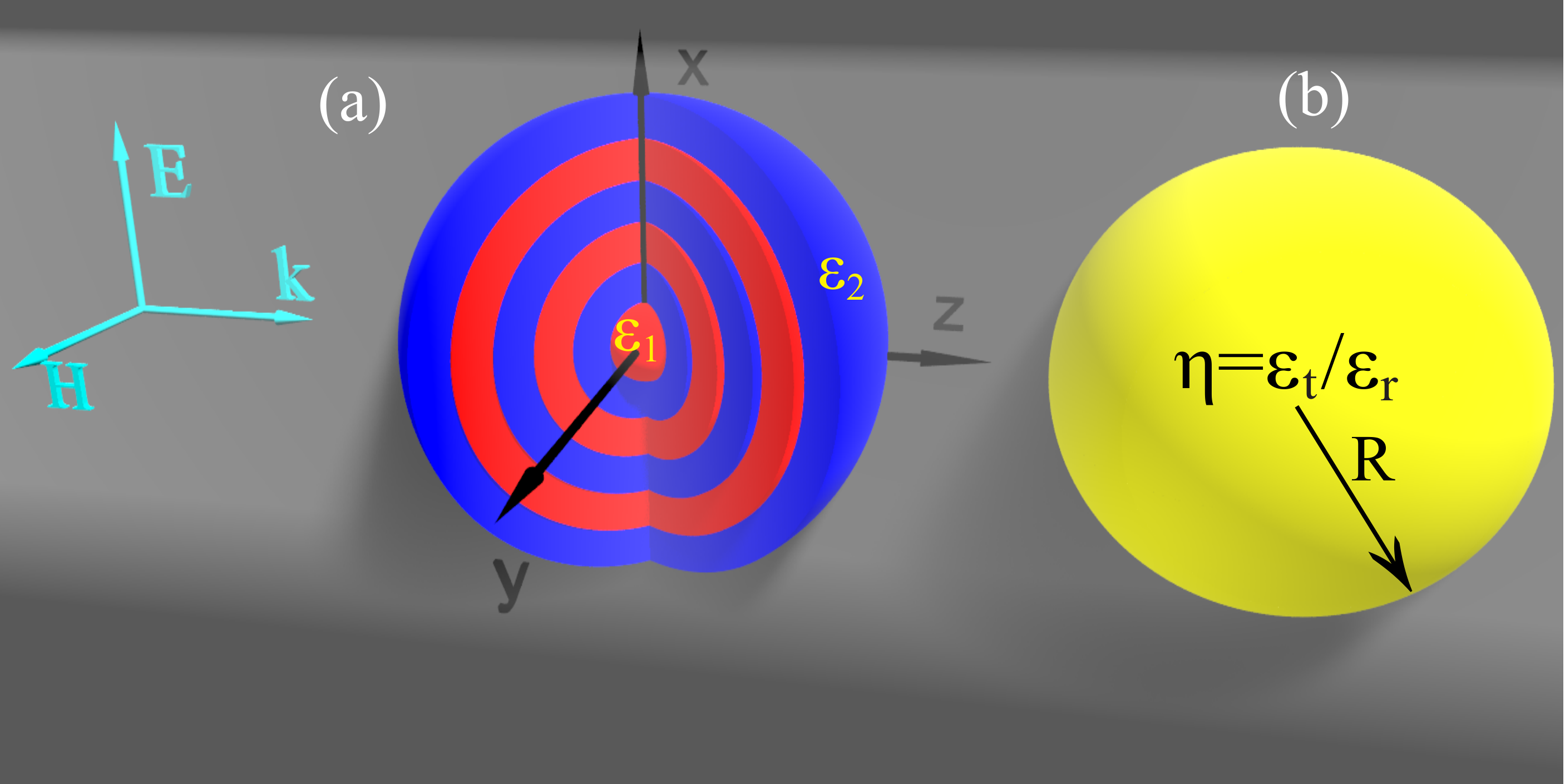}{fig1}{\small (a) Scattering of an $x$ direction polarized (in terms of electric field) plane wave by a multilayered spherical cavity consisting of alternating isotropic layers of permittivities $\varepsilon_1$ and $\varepsilon_2$. The filling factor of the layer of permittivity $\varepsilon_1$ is $f$ in terms of overall layer width. The overall resonator radius is $R$. (b) A homogeneous radially anisotropic sphere of radius $R$ and with radial permittivity $\varepsilon_r$ and transverse permittivity $\varepsilon_t$. The radial anisotropy parameter is defined as $\eta=\varepsilon_t/\varepsilon_r$ and all the materials involved are nonmagnetic.}
%-------------------------------------------------------------------------------

Figure~\ref{fig1}(a) shows the scattering configuration we study: the incident plane wave is polarized along $x$ direction and the multilayered spherical resonator (overall radius $R$) is made of nonmagnetic alternating isotropic layers of two permittivity parameters $\varepsilon_1$  and $\varepsilon_2$ ($\varepsilon_1\neq\varepsilon_2$ ). When each individual layer
width is far smaller then the effective wavelength of the incident wave in the layer, according to the the effective medium theory~\cite{poddubny2013_NP_hyperbolic,liu_q-factor_2016,wu2014_PRX_electrodynamical}, the layered resonator in Fig.~\ref{fig1}(a) can be approximated as a homogeneous radially anisotropic sphere shown in Fig.~\ref{fig1}(b) of the same radius $R$. The corresponding permittivity along radial and transverse directions are respectively:
%==========================================================================
\begin{gather}
\label{effective1}
\varepsilon_r=\varepsilon_1\varepsilon_2/[(1 - f)\varepsilon_1  + f\varepsilon_2],\\
\varepsilon_t=f\varepsilon_1  + (1 - f)\varepsilon_2, %
\end{gather}
%==========================================================================
where $f$ is the filling factor of the $\varepsilon_1$ layer in terms of layer width. The anisotropy parameter $\eta$ is defined as:
%=====================================================================
\begin{equation}
\label{anisotropy_parameter}
\eta=\frac{\varepsilon_t}{\varepsilon_r}. %
\end{equation}
%============================================================================

The scattering of plane waves by radially anisotropic spherical particles (both single layered or multilayered) can be solved analytically through generalized Mie theory~\cite{Bohren1983_book,Qiu2008_JOSAA,Liu2015_OE_Ultra}. The total scattering and absorption cross sections normalized by ${2\pi \over {{k^2}}}$ ($k$ is the angular wave number in the background, which is vacuum in this work) are respectively :
%--------------------------------------------------------------
\begin{gather}
\label{cross_section}
{N_{\rm sca}} = \sum\limits_{n = 1}^\infty  {(2n + 1)({{\left| {{a_n}} \right|}^2}}  + {\left| {{b_n}} \right|^2}),\\
{N_{\rm abs}} = \sum\limits_{n = 1}^\infty  {(2n + 1)[\Upsilon(a_n)+\Upsilon(b_n)}]
\end{gather}
%-------------------------------------------------------------
where the function $\Upsilon(\cdot)$ is defined as $\Upsilon(\cdot)=\rm Re(\cdot)-|\cdot|^2$, and $\rm Re(\cdot)$ means to adopt the real part; $a_n$ and $b_n$ are Mie scattering coefficients, which correspond to electric and magnetic resonance of the $n-th$ order respectively (more specifically $a_1$ and $b_1$ correspond to ED and MD respectively). It is worth mentioning that $b_n$ is dependent on $\varepsilon_t$ only as the magnetic resonances are intrinsically transverse electric and thus are $\varepsilon_r$ independent (thus $b_n$ is not directly dependent on $\eta$); while in contrast $a_n$ is dependent on both $\varepsilon_t$ and $\varepsilon_r$ and thus is directly $\eta$-dependent~\cite{Bohren1983_book,Qiu2008_JOSAA,Liu2015_OE_Ultra}.  Both $a_n$ and $b_n$ can be analytically calculated, and for the simplest case of homogeneous sphere shown in Fig.~\ref{fig1}(b) we have:
%--------------------------------------------------------------
\begin{gather}
%\label{an_bn}
a_n  = {{m\psi _{\tilde n} (m\alpha )\psi '_{n} (\alpha ) - \psi _{n} (\alpha )\psi '_{\tilde n} (m\alpha )} \over {m\psi _{\tilde n} (m\alpha )\xi '_{n} (\alpha ) - \xi _{n} (\alpha )\psi '_{\tilde n} (m\alpha )}} \label{an},\\
b_n  = {{\psi _n (m\alpha )\psi '_n (\alpha ) - m\psi _n (\alpha )\psi '_n (m\alpha )} \over {\psi _n (m\alpha )\xi '_n (\alpha ) - m\xi _n (\alpha )\psi '_n (m\alpha )}} \label{bn},
\end{gather}
%--------------------------------------------------------------
where $\tilde n = \sqrt {n(n + 1)\eta  + {1 \over 4}}  - {1 \over 2}$; $\alpha=kR$; $\psi$ and $\xi$ are Riccati-Bessel functions~\cite{Bohren1983_book}.  When $\eta=1$, $\tilde n=n$ and Eqs.(\ref{an})-(\ref{bn}) will be reduced to the well known expressions of isotropic spheres~\cite{Bohren1983_book}.  According to Eq.(\ref{cross_section}) the normalized scattering cross section contributed by ED and MD are respectively:

%--------------------------------------------------------------
\begin{equation}
\label{cross_section_individual}
{N_{\rm sca}(a_1)} =  {3{{\left| {{a_1}} \right|}^2}},~~~{N_{\rm sca}(b_1)} =  {3{{\left| {{b_1}} \right|}^2}}.
\end{equation}
%-------------------------------------------------------------
At the same time the law of energy conservation requires that~\cite{Ruan2011_APL}:
\begin{equation}
\label{absolute}
|a_n|\le 1,~~~|b_n| \le 1
\end{equation}
which together with Eq. (\ref{cross_section_individual}) immediately requires that the upper limit of the normalized scattering cross section for both ED and MD are $3$. Basically when dipolar resonance are dominantly excited, to be in the superscattering regime requires that ${N_{\rm sca}}>3$. When both $a_n$ and $b_n$ have been obtained, the far-field scattering patterns can be calculated directly~\cite{Bohren1983_book,Liu2015_OE_Ultra}.

\section{Unidirectional superscattering by all-dielectric multilayered cavities of $\eta>1$}

We start with all-dielectric multilayered cavities of $\varepsilon_1,~\varepsilon_2>0$.  It is easy to prove that:
\begin{equation}
\label{anisotropy_1}
\varepsilon_t>\varepsilon_r,~~~\eta>1
\end{equation}
and the highest anisotropy parameter that can be achieved is
\begin{equation}
\label{anisotropy_2}
\eta_{\rm max}=(\varepsilon_1+\varepsilon_2)/2\sqrt{\varepsilon_1\varepsilon_2}
\end{equation}
when $f=1/2$. It is shown that the radial anisotropy can be employed to tune the resonant positions of EDs, which can enable the fully resonant overlapping of EDs and MDs, and result in unidirectional forward superscattering~\cite{Liu2015_OE_Ultra}. For $\eta>1$ though EDs can not be engineered to overlap with MDs of the same mode number (a series of EDs and MDs can be excited at different wavelengths, and a mode number is adopted to differentiate them; for example the resonance excited at the largest wavelength has a mode number $1$~\cite{Liu2015_OE_Ultra}), it is possible to overlap EDs and MDs of different mode numbers.
 
%-------------------------------------------------------------------------------
\pict[0.80]{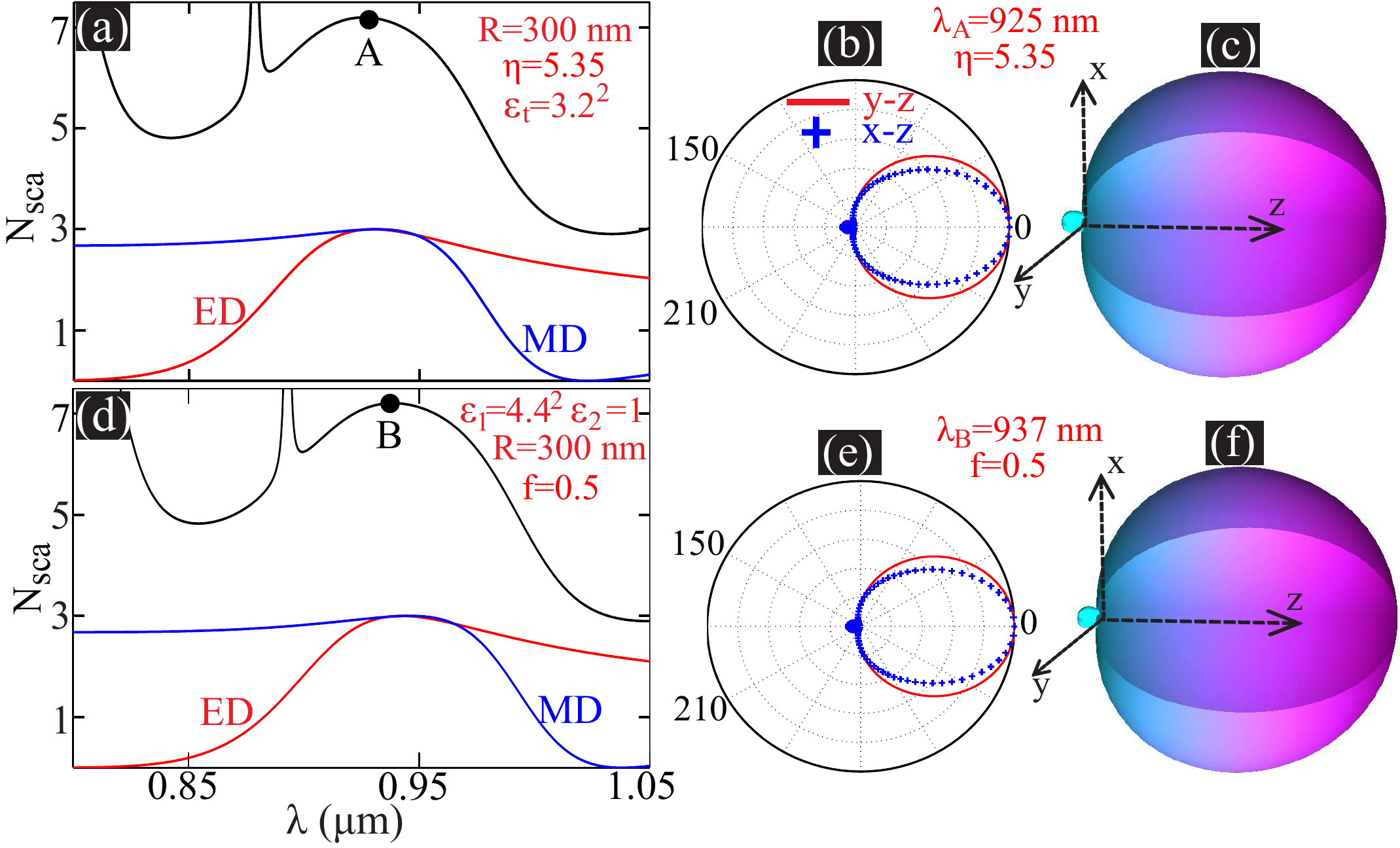}{fig2}{\small (a) Normalized scattering cross section spectra for the homogeneous radially anisotropic sphere, where both total scattering spectra (black curve, as is the case throughout the paper) and those contributed by ED (red curve) and MD (blue curve) are shown. The specific parameters for the sphere are: $R=300$~nm, $\varepsilon_t=3.2^2$ and $\eta=5.35$. The overlapping resonant point is A ($\lambda_A=925$~nm) and at this point the 2D and 3D scattering patterns are shown in (b) (red line and blue crosses indicate angular scattering intensities on the $y-z$ and $x-z$ panes respectively) and (c) respectively. (d) Normalized scattering cross section spectra for the multilayered cavity made up of $30$ layers with alternating $\varepsilon_1=4.4^2$ and $\varepsilon_2=1$ layers of the same width of $10$~nm. The overlapping resonant point is B ($\lambda_B=937$~nm) and at this position the 2D and 3D scattering patterns are shown in (e) and (f) respectively.}
%-------------------------------------------------------------------------------

In Fig.~\ref{fig2}(a) we show the scattering spectra of a homogeneous anisotropic sphere of $R=300$~nm, $\varepsilon_t=3.2^2$ and $\eta=5.35$, where both total scattering spectra (black curve, as is the case throughout the paper) and those contributed by ED and MD are shown. It is clear that ED and MD are resonantly overlapped  (here the mode number for ED and MD are $1$ and $2$ respectively) and the resonant position is indicated by point $A$ of $\lambda_A=925$~nm. At point A it is certainly superscattering as the total scattering is much larger than the single channel limit~\cite{Ruan2011_APL}. Moreover, the overlapping of ED and MD will significantly suppress the backward scattering and enhance the forward scattering, which is shown in Fig.~\ref{fig2}(b) [two-dimensional (2D) scattering patterns on the $x-z$ and $y-z$ planes) and Fig.~\ref{fig2}(c) [full three-dimensional (3D) scattering pattern]. We note here that according to Fig.~\ref{fig2}(b) and (c) the backward scattering has not been fully eliminated and the scattering patterns are not azimuthally symmetric (scattering patterns are not identical on different scattering planes containing $z$ axis).   This is because at point A, despite the resonant excitations of ED and MD, the magnetic quadruple are also present though the magnitude is much smaller (not shown here).  Moreover, at point A the total scattering is larger than sum of those contributed by ED and MD, reconfirming the existence of extra resonances.

The anisotropy parameter discussed above can be realized by an all-dielectric multilayered cavity of $\varepsilon_1=4.4^2$ and $\varepsilon_2=1$, which effectively makes $\varepsilon_t\approx3.2^2$, and $\eta\approx5.35$ when $f=1/2$.  In Fig.~\ref{fig2}(d) we show the scattering spectra of a multilayered cavity consisting of $15$ units: each unit is made up of two layers of the same width of $10$~nm and the permittivity is $\varepsilon_1=4.4^2$ and $\varepsilon_2=1$ respectively, and thus $f=0.5$ and overall radius of the cavity $R=300$~nm. It is clear that the results agree well with those shown in Fig.~\ref{fig2}(a) [the spectra is overall a bit red-shifted, which can be made convergent to those shown in Fig.~\ref{fig2}(a) with decreasing layer width when the effective medium theory is more accurate], justifying the validity of the effective medium theory. The resonant overlapping position is indicated by point B of $\lambda_B=937$~nm, and the corresponding scattering patterns are shown in Fig.~\ref{fig2}(e)-(f), which also agree well with those obtained through effective medium theory shown in Fig.~\ref{fig2}(b)-(c), and confirms the unidirectional superscattering of the all-dielectric multilayered cavity.

\section{Unidirectional superscattering by plasmonic multilayered cavities of $\eta<1$}

%-------------------------------------------------------------------------------
\pict[0.80]{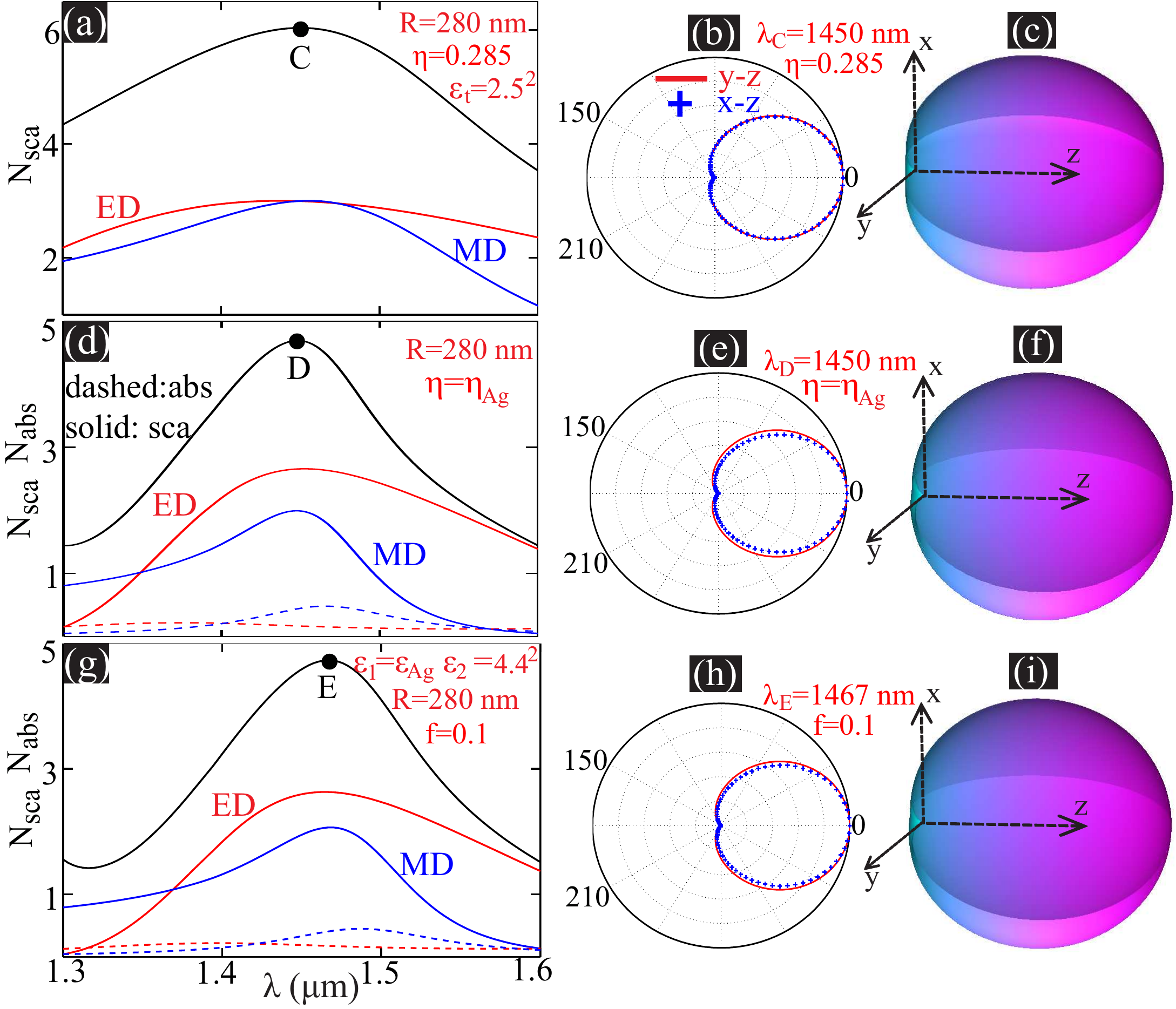}{fig3}{\small (a) Normalized scattering cross section spectra for a homogeneous radially anisotropic sphere: $R=280$~nm, $\varepsilon_t=2.5^2$ and $\eta=0.285$. The overlapping resonant point is C ($\lambda_C=1450$~nm) and at this point the 2D and 3D scattering patterns are shown in (b) and (c) respectively. (d) Normalized scattering (solid curves) and absorption (ED: red dashed; MD: blue dashed) cross section spectra for a homogeneous radially anisotropic sphere of $R=300$~nm, for which the effective parameters are obtained through Eq.~(\ref{effective1})-(\ref{anisotropy_parameter}) with $\varepsilon_1=\varepsilon_{Ag}$, $\varepsilon_2=4.4^2$ and $f=0.1$. The dispersive anisotropy parameter obtained is $\eta=\eta_{\rm Ag}$. The overlapping resonant point is D ($\lambda_D=1450$~nm) and the scattering patterns are shown in (e) and (f). (g) The scattering and absorption spectra for the multilayered cavity made up of $56$ layers with alternating silver and dielectric ($\varepsilon_1=4.4^2$) layers of width $1$~nm and $9$~nm respectively. The overlapping resonant point is E ($\lambda_E=1467$~nm) and the scattering patterns are shown in (h) and (i).}
%-------------------------------------------------------------------------------

To achieve unidirectional superscattering relying on overlapping EDs and MDs of the same mode number requires $\eta<1$.  In Fig.~\ref{fig3}(a) we show the scattering spectra of a homogeneous anisotropic sphere of $R=280$~nm, $\varepsilon_t=2.5^2$ and $\eta=0.285$, where ED and MD of the same mode number of $1$ (ED and MD excited at the largest wavelength) are resonantly overlapped and the resonant position is $C$ ($\lambda_C=1450$~nm). At this point it is unidirectional superscattering as proved by the scattering patterns shown in Fig.~\ref{fig3}(b)-(c). 

The condition of $\eta<1$  can be satisfied in hybrid plasmonic-dielectric multilayered cavities of $\varepsilon_1<0$ and $\varepsilon_2>0$ according to Eq.~(\ref{effective1})-(\ref{anisotropy_parameter}). For a multilayered cavity of $\varepsilon_t=2.5^2$ and $\eta=0.285$, it requires that $\varepsilon_1\approx-112$ and $f=0.1$ if $\varepsilon_2=4.4^2$. We can design a silver-dielectric ($\varepsilon_2=4.4^2$) hybrid cavity of $f=0.1$ and $R=280$~nm, where the permittivity of silver ($\varepsilon_{Ag}$) is taken from experimental data~\cite{Johnson1972_PRB}. Of such a plasmonic cavity though the effective parameters ($\varepsilon_t$, $\varepsilon_r$ and $\eta_{\rm Ag}$) are dispersive, the resonant overlapping of ED and MD poses only the requirement that the anisotropy parameter of $\eta=0.285$ is provided at the resonant position, which silver can meet [$Re(\varepsilon_{Ag})\approx-112$ when $\lambda=1450$~nm]. In Fig.~\ref{fig3}(d) we show both the scattering and absorption spectra of a homogeneous sphere of effective parameters obtained through Eq.~(\ref{effective1})-(\ref{anisotropy_parameter}) with $\varepsilon_1=\varepsilon_{Ag}$, $\varepsilon_2=4.4^2$ and $f=0.1$. The resonant position is $D$ ($\lambda_D=1450$~nm), where the ED and MD are resonantly overlapped. In contrast to the results in Fig.~\ref{fig3}(a), the ED and MD are not of the same scattering magnitude due to the effect of Ohmic losses of silver, which lead to different absorptions of ED and MD (see the dashed absorption curves). Nevertheless, as is shown in Fig.~\ref{fig3}(e)-(f), unidirectional superscattering with negligible backward scattering can still be achieved though the pattern is not rigourously azimuthally symmetric.

Now we turn to the multilayered plasmonic cavity consisting of $28$ units: each unit is made up of a silver layer of width $1$~nm and a isotropic dielectric layer ($\varepsilon_1=4.4^2$) of width $9$~nm, and thus $f=0.1$ and overall radius of the cavity $R=280$~nm. In Fig.~\ref{fig3}(g) we show the scattering and absorption spectra of such a multilayered cavity. The results agree well with those shown in Fig.~\ref{fig3}(d) obtained through effective medium theory. The resonant overlapping position is indicated by point E of $\lambda_E=1467$~nm, and the corresponding scattering patterns are shown in Fig.~\ref{fig3}(h)-(i), which confirms the unidirectional superscattering of the multilayered plasmonic cavity. It is worth mentioning that here we neglect the nonlocal effect of the thin silver layer~\cite{Boardman1982_book}. Even when the nonlocal effect is present and not negligible, the specific anisotropy parameter is only required at a single resonant wavelength, which can still be satisfied through geometric tuning.

%-------------------------------------------------------------------------------
\pict[0.65]{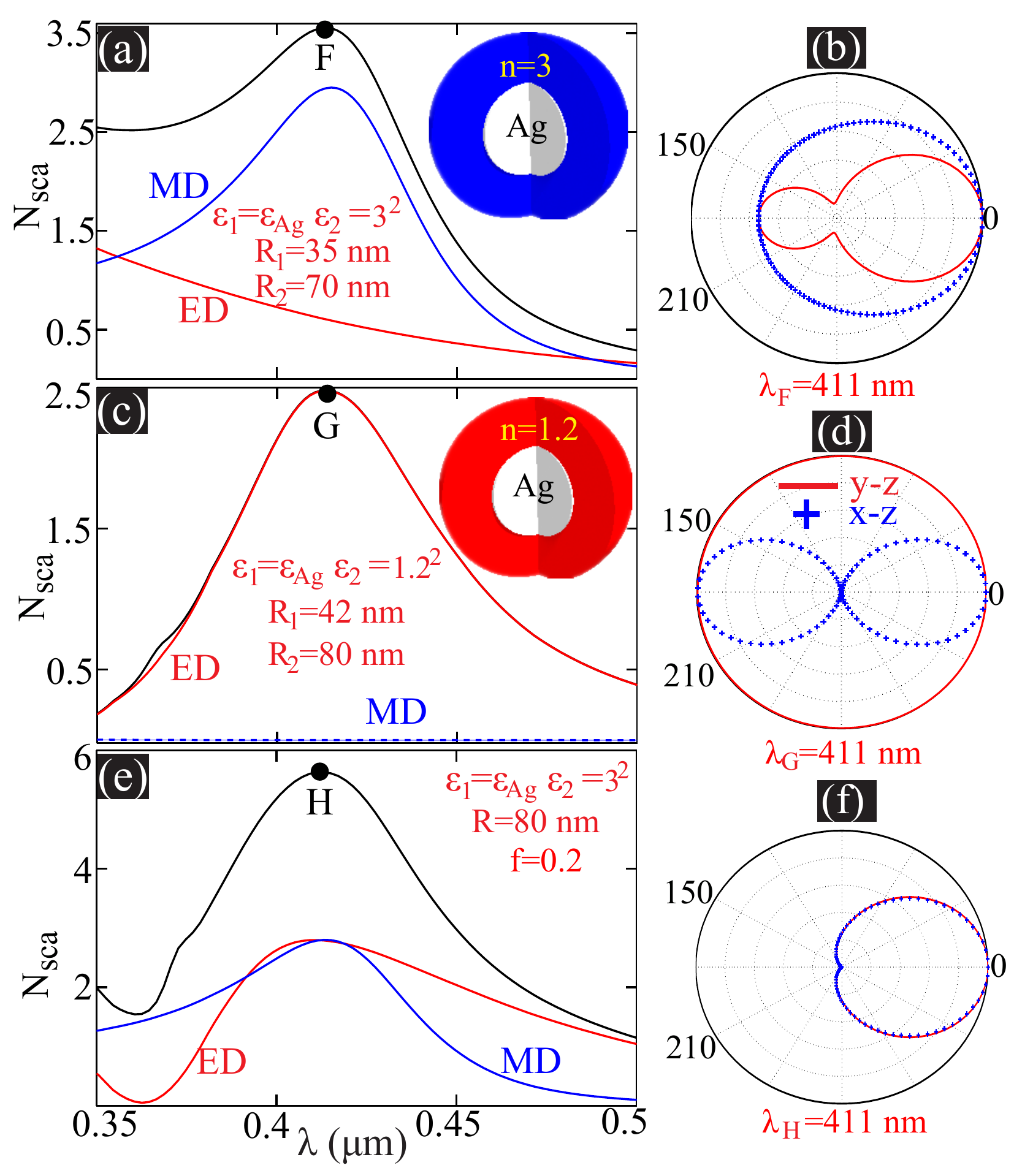}{fig4}{\small Normalized scattering cross section spectra of two layered silver-dielectric cavity of dielectric permittivity $\varepsilon_2=3$, $R_1=35$~nm, $R_2=70$~nm in (a)  and  $\varepsilon_2=1.2$, $R_1=42$~nm, $R_2=80$~nm in (b). The resonant positions are indicated by F and G ($\lambda_F=\lambda_G=411$~nm), and at those points the 2D scattering patterns are shown in (b) and (d) respectively. (e) Normalized scattering cross section  for the multilayered cavity made up of $32$ layers with alternating silver and dielectric ($\varepsilon_2=3^2$) layers of width $1$~nm and $5$~nm respectively. The overlapping resonant point is H ($\lambda_H=411$~nm) and the 2D scattering patterns are shown in (f).}
%-------------------------------------------------------------------------------

It is well known that in a simple two layered core-shell metal-dielectric cavity, the ED and MD can be tuned to resonantly overlap, producing unidirectional forward superscattering~\cite{Liu2012_ACSNANO}. Now the question is: why do we need the many layered plasmonic cavity described above to overlap EDs and MDs?  The problem is that though ED and MD can resonantly overlap at relatively large wavelength, it is not possible to excite and overlap them at wavelengths close to the plasmon frequency  in a two layered plasmonic resonator. To efficiently excite MDs at small wavelengths, it is known that high permittivity dielectric materials are required~\cite{Kuznetsov2012_SciRep,Evlyukhin2012_NL}. While high permittivity dielectric layer will red-shift the EDs, making them spectrally separated from MDs. In Fig.~\ref{fig4}(a) we show the scattering spectra of a silver-dielectric ($\varepsilon_1=3^2$) core-shell cavity (inner radius $R_1=35$~nm and outer radius $R_2=70$~nm) where the MD resonates at point F ($\lambda_F=411$~nm). However at this point the scattering magnitude of ED is much smaller, rendering significant backward scattering, as can be observed in the scattering patterns shown in Fig.~\ref{fig4}(b). At the same time, with a low index dielectric layer, the EDs can be efficiently excited at small wavelength, while under such circumstance the MD can not be efficiently excited, making the scattering pattern a typical ED type with equal forward and backward scattering. This is demonstrated in Fig.~\ref{fig4}(c) where we show the scattering spectra of a silver-dielectric ($\varepsilon_1=1.2^2$) core-shell resonator ($R_1=42$~nm, $R_2=80$~nm). Here the ED resonates at point G ($\lambda_G=411$~nm), where however the MD excitation is negligible. As a result, the scattering is contributed only by ED and thus not unidirectional [see the scattering patterns shown in Fig.~\ref{fig4}(d)].

The challenge of overlapping EDs and MDs at small wavelengths close to the plasmon frequency is not insurmountable for multilayered plasmonic cavities of effective radial anisotropy.    we employ a multilayered plasmonic cavity consisting of $16$ units: each unit is made up of a silver layer of width $1$~nm and a isotropic dielectric layer ($\varepsilon_1=3^2$) of width $4$~nm, and thus $f=0.2$ and overall radius of the cavity $R=80$~nm. In Fig.~\ref{fig4}(e) we show the scattering spectra of such a plasmonic multilayered cavity. It is clear that ED and MD can be made to resonantly overlap at point H ($\lambda_H=411$~nm), which induces unidirectional forward superscattering [see the scattering patterns shown in Fig.~\ref{fig4}(f)].

\section{Conclusions and discussions}

To conclude, in this work we study the plane wave scattering by multilayered cavities that possess large effective radial anisotropy. In such cavities, relying on the effective radial anisotropy, the EDs and MDs can be tuned to resonantly overlap,  which thus satisfies the Kerker's condition and produces unidirectional forward superscattering. It is demonstrated that such scattering shaping can be realized in both all-dielectric and plasmonic multilayered cavities. Moreover it is shown that in plasmonic cavities of effective radial anisotropy, the EDs and MDs can be made to resonantly overlap at small wavelengths in the violet spectral regime, which is not accessible for simple two layered metal-dielectric core-shell resonators. We note here that in this work we have confined our studies to dipolar modes, and based on the same approach higher order modes can be made to overlap, which can produced more collimated forward superscattering~\cite{Liu2015_OE_Ultra}. Also, we have studied only the case of $\eta>0$, and quite naturally hyperbolic cavities of $\eta<0$ can be studied using the same method~\cite{poddubny2013_NP_hyperbolic,wu2014_PRX_electrodynamical}. Moreover, the principle we have revealed is general, which is also applicable to resonators of other shapes and to other kinds of anionotropy such as magnetic anisotropy or the intrinsic huge anisotropy of 2D materials. Such mechanism of resonance control based on effective anisotropy can shed new light to many particle scattering problems and is quite promising for various applications in the fields of nanoantennas, solar cells, bio-sensing and so on.

%{\em Acknowledgements:}
 \section*{Acknowledgments}

 We thank Andrey E. Miroshnichenko and Yuri S. Kivshar for valuable discussions.  We also acknowledge the financial support from the National Natural Science Foundation of China (Grant number: $11404403$) and the Basic Research Scheme of College of Optoelectronic Science and Engineering, National University of Defense Technology.

\end{sloppy}
\end{document}